# Non-Unitary Quantum Walks on Hyper-Cycles

Dmitry Solenov and Leonid Fedichkin

Center for Quantum Device Technology,
Department of Physics, Clarkson University, Potsdam, NY 13699-5721
Electronic addresses: solenov@clarkson.edu, leonid@clarkson.edu

PACS: 03.65.Xp, 03.67.Lx, 05.40.Fb, 05.45.Mt                              September 2005

We present analytical treatment of quantum walks on multidimensional hyper-cycle graphs. We derive the analytical expression of the probability distribution for strong and weak decoherence regimes. Upper bound to mixing time is obtained.

## I. Introduction

Significant advances in nanotechnology have stimulated development of novel approaches to harness quantum-mechanical properties in computation. The milestones on this way were Shor's factoring [1] and Grover's search [2] algorithms, which have marked the appearance of the large family of discrete-time quantum computing algorithms [3]. Recently, a new promising idea—quantum walks has become a subject of broad interest in quantum computation community and beyond [4-17]. Since they were brought to attention [18], quantum random walks have been studied as both discrete-time [19-21] and continuous-time walks [22-25]. Appearing in many publications, discrete-time quantum walks have been argued to have strong ties with the traditional algorithmic approach and require a full-scale quantum computer to be built. The latter task in its various aspects has obtained significant attention throughout the research community, but is yet to be accomplished. At the same time, there is a hope that continuous-time quantum walks can potentially outstrip the event, relaxing the requirement, as they resemble many natural quantum-mechanical systems.

The interest in quantum walks has been heated up by the enduring success of their predecessor—classical random walks [26] and by the emergence of evidence that the former can outperform its classical counterpart even in the presence of decoherence. In its quantum version the walking particle is assigned certain amplitude of a jump between two connected vertexes of the graph that defines the structure. In addition the measurement is performed on each vertex which introduces non-unitary dynamics. In most of the investigations the non-unitarity comes through an introduction of probable vertex readout [22,25]. In continuous-time quantum walks it is more natural to account for weak continuous observation of the system by its environment, since this interaction inevitably enters in any practical application. For the cycle graphs, see Fig.1, it was shown that introduction of interaction with the heat bath through point contact scenario [27-29], when each node is constantly monitored by an attached point contact, leads to the same set of equations with naturally arising probability of vertex measurement [30].

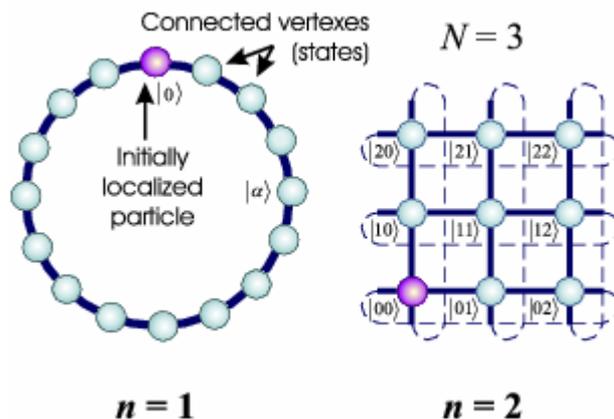

Figure 1. Schematic construction of $n$-dimensional hyper-cycles. One- and two- dimensional cases are shown.

One of the important properties of the walks that one wants to exploit in various algorithmic solutions [31-38] is mixing. The notion of mixing came from the classical random walks where one can define the "mixing time" at which the distribution of probability happens to be sufficiently close to the stationary (often uniform) distribution. The analysis of mixing time in quantum walks reveals two types of mixing. The first one, instantaneous mixing, is defined as the time at which the probability is close to the uniform one but can diverge rapidly a moment latter. This type of mixing is determined by the structure of the graph and often may not exist. The other, average mixing, was introduced to handle the situations when the probability does not evolve to any stationary distribution. Introduction of decoherence as a continuous measurement introduces "natural averaging" resolving the latter problem. Interestingly to note, the presence of decoherence in the case of quantum walks does not simply fall into the paradigm of a certain threshold [39], which happens with conventional quantum computing, instead, it enhances the mixing, reducing the mixing time, as was shown numerically for discrete-time walks on the cycle graphs by Kendon and Tregenna [40]. The mixing time there was argued to have linear dependence on the size of the graph, as opposed to quadratic dependence in classical random walks [26]. There were also analytical evidences of the latter for continuous-time case [30].

Despite the recent intense interest in the continuous-time quantum walks many of its properties are still not well understood and require careful investigation. The present paper contributes to the study of the decoherence effect on the hyper-cycle graphs. Namely, we discuss continuous-time quantum walks on $n$-dimensional hyper-cycles with naturally arising decoherence. Starting up with the review of decoherence-affected walks on a single cycle, see Section II, we generalize the system to $n$ dimensions in Section III. It is shown that the solution to the density matrix in $n$ dimensions is based on the dynamics of a single cycle, as it was demonstrated earlier for $n$-dimensional hypercube [22,25]. We obtain also expressions for the upper bound to mixing time in both quantum and classical limits, see Section IV.

## II. The Cycle Graph

Quantum walks on the cycle graphs in the presence of decoherence has been studied numerically in its discrete-time interpretation [40]. It has appeared that small amount of decoherence can actually enhance mixing property in such the system. Resent analytical studies of continues-time walks made by the authors [30] have shown similar behavior. In order to proceed with complex multidimensional system, we cite important results of single-cycle studies made for weak decoherence (quantum) limit. We also obtain analytical solution for strong decoherence scenario (classical limit), as it will be useful for further analysis.

Continuous-time quantum walks on a cycle graph of size $N$ we consider are defined [30] by the following equation for the density matrix elements

$$\frac{d}{dt}\rho_{\alpha\beta} = \frac{i}{4}\left(\rho_{\alpha,\beta+1} - \rho_{\alpha+1,\beta} - \rho_{\alpha-1,\beta} + \rho_{\alpha,\beta-1}\right) - \Gamma\left(1 - \delta_{\alpha\beta}\right)\rho_{\alpha\beta}. \quad (2.1)$$

Here indexes $\alpha, \beta$ (as well as $\mu, \nu$ latter in the text) number the electron states, $|\alpha\rangle$, on each vertex, and run from $0$ to $N-1$. Equation (2.1) also assumes cyclic substitution of indexes on the right hand-side. It is convenient to utilize vector notations for the density matrix, introducing

$$V(t) \equiv \sum_{\mu\nu} \rho_{\mu\nu}(t)|\mu\rangle \otimes |\nu\rangle. \quad (2.2)$$

Considering the above definition, system (2.1) can be formulated in a vector-matrix form,

$$\frac{d}{dt}V = -i(H \otimes I - I \otimes H)V - \Gamma\left(I \otimes I - \sum_{\mu} P_\mu \otimes P_\mu\right)V, \quad (2.3)$$

where $I$ is an identity matrix of size $N \times N$, $H$ is the Hamiltonian of the coherent walks, i.e.

$$H = \frac{1}{4}\sum_{\mu}\left(|\mu\rangle\langle\mu+1| + |\mu+1\rangle\langle\mu|\right). \quad (2.4)$$

Here, the set of projectors is define as

$$P_\mu = |\mu\rangle\langle\mu|. \tag{2.5}$$

The latter, in fact, represents measurements performed on each node of the cycle. The convenience of (2.3) is that it immediately gives formal solution to the density matrix (vector) in the form that will be helpful in multidimensional analysis,

$$V(t) = M(t)V(0), \tag{2.6}$$

where the "propagation matrix" $M(t)$ is

$$M(t) = e^{it(I\otimes H - H\otimes I)V - \Gamma t\left(I\otimes I - \sum_\mu P_\mu \otimes P_\mu\right)}. \tag{2.7}$$

Obtaining explicit solution to (2.1) and (2.7) is a tough problem in itself. It splits into several modes depending on the strength of decoherence introduced. In particular, one can analytically study weak and strong decoherence limits by allowing $\Gamma \ll 1$ and $\Gamma \gg 1$. The intermediate dissipation is quite hard to handle analytically. In this case, numerical simulations, which go beyond the scope of this paper, may be useful.

Let us first, cite the result of Ref. [30] obtained in the weak decoherence limit. For the classical initial distribution $\rho_{\alpha\beta}(0) = \delta_{\alpha,0}\delta_{\beta,0}$ the density matrix, which for this special case will be denoted as $C_{\alpha\beta}(t)$, is

$$C_{\alpha\beta}(t) \stackrel{\Gamma\ll 1}{=} \frac{\delta_{\alpha\beta}}{N} + i^{\alpha-\beta}\sum_{m,n=0}^{N-1}\frac{1-\delta_{m+n,0}-\delta_{m+n,N}}{N^2} \times$$

$$\times \left[\delta_{mn}e^{-\Gamma\frac{N-1}{N}t} + (1-\delta_{mn})e^{-\Gamma\frac{N-2}{N}t}\right]e^{it\sin\frac{\pi(m+n)}{N}\cos\frac{\pi(m-n)}{N}+\frac{2\pi i}{N}(m\alpha+n\beta)}. \tag{2.8}$$

To obtain the approximation to the density matrix in the limit of large $\Gamma$, we note that to the first order in $1/\Gamma$ only the main and the two adjacent diagonals of the density matrix are important. Indeed, strong measurement of each vertex works toward destruction of entanglement between states on different nodes. The interaction between the nodes, which otherwise create the entangled states, is suppressed by measurements on each node which results in Zeno-like behavior. The effect is magnified for the vertexes that are not immediate neighbors since the measurements are performed on each intermediate vertex. As a result, the density matrix elements significantly decrease in magnitude as they depart from the main diagonal. The trend can be also obtained formally analyzing system (2.1). One can show that the density matrix elements on diagonals $d_\alpha^j(t) \equiv i^j \rho_{\alpha+j,\alpha}(t)$ obey the relation $d_\alpha^{j+1}(t) \propto \frac{1}{\Gamma}O(d_\beta^j(t))$ for any $\alpha, \beta$ and for $j$ from $0$ to $N-2$. As far as one is interested in the solution to the first order in $1/\Gamma$, only the main- and the two adjacent sub-diagonals, i.e. $j = 0,1$, are retained. The corresponding equations are

$$\frac{d}{dt}d_\alpha^0(t) = \frac{1}{2}\left[d_\alpha^1(t) - d_{\alpha-1}^1(t)\right] \quad \text{and} \quad \frac{d}{dt}d_\alpha^1(t) = \frac{1}{2}\left[d_{\alpha+1}^0(t) - d_\alpha^0(t)\right] - \Gamma d_\alpha^1(t). \tag{2.9}$$

The system (2.9) has the solution

$$d_\alpha^0(t) = \frac{1}{N}\sum_{k=0}^{N-1} e^{\frac{2\pi i k\alpha}{N} - \frac{t}{\Gamma}\sin^2\left(\frac{\pi k}{N}\right)} \tag{2.10a}$$

and

$$d_\alpha^1(t) = \frac{1}{N}\sum_{k=0}^{N-1}\frac{i\sin\pi k/N}{\Gamma}\left[e^{-\Gamma t} - e^{-\frac{t}{\Gamma}\sin^2(\pi k/N)}\right]e^{\frac{i\pi k(2\alpha+1)}{N}}, \tag{2.10b}$$

Here we have considered $d_\alpha^j(0) = \delta_{\alpha,0}\delta_{j,0}$ as an initial condition, and have shown terms up to order $1/\Gamma$. The expression for the density matrix is

$$C_{\alpha\beta}(t) \equiv \rho_{\alpha\beta}(t) \stackrel{\Gamma \gg 1}{=} \delta_{\alpha\beta}d_\alpha^0(t) + \delta_{\alpha+1,\beta}d_\alpha^1(t) + \delta_{\alpha,\beta+1}d_\beta^1(t) + O(1/\Gamma^2). \tag{2.11}$$

In general, the dynamics on the cycle is given by the propagation matrix

$$M_{(\alpha\beta)(\mu\nu)}(t) = \delta_{\mu\nu}C_{\alpha-\mu,\beta-\mu}(t) + (1-\delta_{\mu\nu})Q_{\alpha-\mu,\beta-\mu}(t), \tag{2.12}$$

where $C_{\alpha,\beta}(t)$-term, considered as (2.8) or (2.11) depending on the magnitude of $\Gamma$, is accountable for classical (unentangled) initial distribution, while the last term is only necessary when we deal with non-classical initial distribution. In many practical applications and experimental studies initial distribution is classical and we can set $Q_{\alpha,\beta}(t) \to 0$. Note that in this case the probability distribution at time $t$ is

$$P_\alpha(t) = \sum_\mu C_{\alpha-\mu,\beta-\mu}(t)\rho_{\mu\mu}(0). \tag{2.13}$$

### III. Multidimensional Cycles

As we have formulated the convenient toolkit on a single cycle problem, we can proceed with incorporating higher dimensionality. Generally, hyper-cycle of $n$ dimensions and size $N$ is build by cloning the system of $n-1$ dimensions $N$ times and connecting the corresponding vertexes, see Fig.1, so that the number of vertexes becomes $N^n$. As one can see from the figure, each vertex is now defined by the $n$-digit number of base $N$, so is the state corresponding to it. The density matrix in its vector form can now be formulated as

$$\mathbf{V}(t) \equiv \sum_{\boldsymbol{\mu\nu}} \boldsymbol{\rho_{\mu\nu}}(t)|\boldsymbol{\mu}\rangle \otimes |\boldsymbol{\nu}\rangle, \tag{3.1}$$

where $\boldsymbol{\mu} = \mu_1\mu_2...\mu_{n-1}$ and each digit $\mu_j$ runs from 0 to $N-1$. From now on we use bold-face symbols to refer to hyper-cycle related structures. The Hamiltonian of the coherent evolution is

$$\mathbf{H} = \frac{1}{n}\sum_{j=1}^n I \otimes ... \otimes \underbrace{H}_{j^{\text{th}}} \otimes ... \otimes I. \tag{3.2}$$

Here we have introduced $1/n$ factor to keep the energy of the overall system the same.

While expanding the graph, we still measure each of its vertexes individually, which introduces a set of projectors that operate separately on each vertex. The latter fact suggest simple generalization of (2.5)

$$\mathbf{P}_\mu^j = \frac{1}{n} I \otimes ... \otimes \underbrace{P_\mu}_{j^{\text{th}}} \otimes ... \otimes I, \tag{3.3}$$

where $1/n$ factor is introduced to keep proper completeness relation. The electron dynamics on the hyper-cycle graph is given now by the equation

$$\frac{d}{dt}\mathbf{V} = (i\mathbf{I} \otimes \mathbf{H} - i\mathbf{H} \otimes \mathbf{I})\mathbf{V} - \Gamma\left(\mathbf{I} \otimes \mathbf{I} - \sum_{\mu,j}\mathbf{P}_\mu^j \otimes \mathbf{P}_\mu^j\right)\mathbf{V}. \tag{3.4}$$

Here $\mathbf{I} = \underbrace{I \otimes ... \otimes I}_{n}$. Following the technique of [25] proposed for hyper-cube calculations, one can demonstrate that the solution to hyper-cycles is formed by the one of a single cycle (2.12). To show this let us write each term in (3.4) explicitly

$$\mathbf{I} \otimes \mathbf{H} = \frac{1}{n}\sum_{j=1}^{n}(I \otimes I) \otimes ... \otimes \underbrace{(I \otimes H)}_{j^{\text{th}}} \otimes ... \otimes (I \otimes I), \tag{3.5}$$

$$\mathbf{H} \otimes \mathbf{I} = \frac{1}{n}\sum_{j=1}^{n}(I \otimes I) \otimes ... \otimes \underbrace{(H \otimes I)}_{j^{\text{th}}} \otimes ... \otimes (I \otimes I) \tag{3.6}$$

and

$$\sum_{\mu,j}\mathbf{P}_\mu^j \otimes \mathbf{P}_\mu^j = \frac{1}{n}\sum_{j}(I \otimes I) \otimes ... \otimes \underbrace{\left(\sum_\mu P_\mu \otimes P_\mu\right)}_{j^{\text{th}}} \otimes ... \otimes (I \otimes I). \tag{3.7}$$

Substituting these expressions back into (3.4) and integrating the equation one obtains the propagation matrix

$$\mathbf{M}(t) = e^{\sum_j (I \otimes I) \otimes ... A ... \otimes (I \otimes I)} = \left(e^A\right)^{\otimes n}, \tag{3.8}$$

where

$$A \equiv \frac{t}{n}\left[(I \otimes iH - iH \otimes I) - \Gamma\left(I \otimes I - \sum_\mu P_\mu \otimes P_\mu\right)\right]. \tag{3.9}$$

The exponential of $A$ in (3.8) is a familiar object, see (2.7). It is equal to $M(t/n)$. The density matrix of the system is now easy to construct. One yields

$$\boldsymbol{\rho}_{\alpha\beta}(t) = \sum_{\boldsymbol{\mu\nu}}\left\{\prod_{j=0}^{n-1}M_{(\alpha_j\beta_j)(\mu_j\nu_j)}\right\}\boldsymbol{\rho}_{\boldsymbol{\mu\nu}}(0). \tag{3.10}$$

The resulting expression for the probability distribution in the case of classical initial conditions is

$$\mathbf{P}_{\boldsymbol{\alpha}}(t) = \sum_{\boldsymbol{\mu}}\left\{\prod_{j=0}^{n-1}C_{\alpha_j-\mu_j,\alpha_j-\mu_j}(t/n)\right\}\boldsymbol{\rho}_{\boldsymbol{\mu\mu}}(0), \tag{3.11}$$

where $C_{\alpha,\beta}(t)$ is given by (2.8) or (2.11).

On Fig.2 we plot the probability distribution for the walks on 3-dimensional hyper-cycle of size $N = 3$, placing the walking particle initially at node $00$. The evolution for small and large $\Gamma$ is shown. In the first case, Fig.2A, the probability spreads around, developing symmetrical oscillations pattern, which converges to the uniform distribution later on. One observes the concentration of the beats on $000$, $111$ and $222$ vertexes of the graph. The case $\Gamma \gg 1$, presented on Fig.2B, shows completely different behavior. The density diffuses slowly across the structure, reproducing classical diffusion scenario. Continuous-time quantum walks, as we see, present an interplay of two opposite phenomenon—wave propagation and diffusion. In the next section we will show how their competition may potentially lead to non-trivial dependence for the characteristics like mixing time.

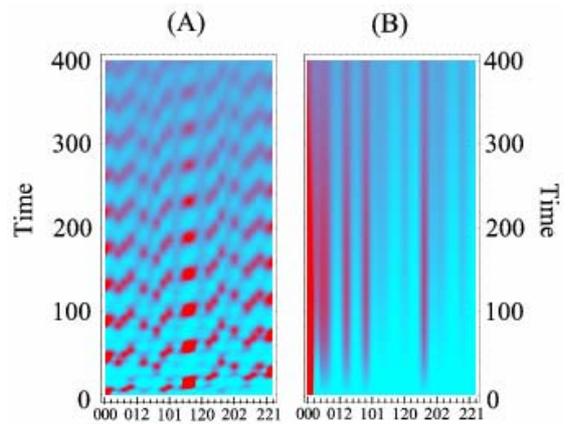

Figure 2. The probability distribution for 3-dimensional hyper-cycle of size $N = 3$. Quantum mode (A) for $\Gamma = 0.05$, and classical mode (B) for $\Gamma = 50$ are shown. The vertexes are referred by 3-digit numbers of base $n = 3$ as mentioned in the text.

## IV. The Mixing Time

One of the important properties of quantum walks is their potential ability to fast mixing. Indeed, almost instantaneous onset of the interference pattern that covers all the graph, as shown on Fig.2a, seems promising. The important question, though, is how mixing time depends on the graph size and dimensionality throughout different degrees of decoherence. Obtaining analytical expressions to mixing time even for relatively simple one-dimensional cycle systems is difficult. In order to analyze classical-vs.-quantum modes for mixing on hyper-cycle structures we obtain upper-bound estimates for $\Gamma \ll 1$ and $\Gamma \gg 1$ cases based on the solution (3.11).

As usual [30], we place the walking particle at node $\mu_j = 0$, i.e.

$$\rho_{\mu\mu}(0) = \prod_{j=0}^{n-1} \delta_{\mu_j,0} \qquad (4.1)$$

and allow it to evolve in accordance with (3.4). The mixing time, $t_{mix}$, is then defined as the minimum time $t_m$ at which the probability distribution is sufficiently close to uniform [30,40]

$$\sum_{\boldsymbol{\alpha}} \left| \mathbf{P}_{\boldsymbol{\alpha}}(t_m) - \frac{1}{N^n} \right| \leq \varepsilon . \qquad (4.2)$$

Here $\varepsilon$ defines the degree of mixing. Substituting (3.12) into (4.2) and considering (4.1), after some algebra, see Appendix A for details, one obtains

$$\sum_{\alpha} \left| P_\alpha(t_m/n) - \frac{1}{N} \right| \leq \frac{\varepsilon/n}{1 + N^n \varepsilon} . \qquad (4.3)$$

Expression (4.3) is a familiar inequality. It coincides with the definition of the mixing time of a single isolated cycle up to the time scaling and some replacement for $\varepsilon$, i.e. $t_m \to t_m/n$ and $\varepsilon \to \frac{\varepsilon/n}{1 + N^n \varepsilon}$. Note that in expression (4.3) we have made no assumptions regarding the structure of the probability distribution, except for the initial condition in the form (4.1).

In the limit of small decoherence, $\Gamma \ll 1$, the mixing time of a single cycle was found [30] to be bounded by $N \ln(4/\varepsilon)/\Gamma$. Later investigations have shown that even a tighter bound can be obtained [41], namely

$$t_{mix}^{(1)} \leq \frac{1}{\Gamma} \frac{N}{N-2} \ln\left(\frac{N+1}{\varepsilon}\right). \qquad (4.4)$$

Taking (4.4) into account, we immediately find the upper-bound to mixing time for *n*-dimensional cycle, $t_{mix}^{(n)}$, as

$$t_{mix}^{(n)} \leq \frac{n}{\Gamma} \frac{N}{N-1} \ln\left[\frac{n(N+1)(1+\varepsilon N^n)}{\varepsilon}\right]. \qquad (4.5)$$

From (4.5) it follows that increase of dimensionality introduces $n \log n$-dependence for the bound to mixing time as far as one keeps $\varepsilon N^n \leq 1$.

The strong decoherence scenario requires the analysis of (4.3) with the probability distribution defined by (2.11). Leaving the detailed derivations in Appendix B, we show the resulting expression for the mixing time,

$$t_{mix}^{(1)} \leq \frac{\Gamma N^2}{4} \ln\left(\frac{2+\varepsilon}{\varepsilon}\right). \qquad (4.6)$$

In $n$ dimensions this results in the following upper-bound estimate

$$t_{mix}^{(n)} \leq \frac{\Gamma N^2 n}{4} \ln\left[\frac{n(2+\varepsilon)(1+\varepsilon N^n)}{\varepsilon}\right]. \tag{4.7}$$

The above analysis of hyper-cycle graphs shows that the mixing time upper-bound increases by the factor of $N$ as one goes from quantum to classical limits. Another interesting feature coming out from expressions (4.4-4.7) is their dependence on $\Gamma$. One can notice that for any valid $n$, there must be an optimal $\Gamma$-interval at which mixing time is minimized.

We should note also, that the above analysis is performed to estimate non-instantaneous mixing time, i.e. the time from which the walking particle becomes $\varepsilon$-uniform distributed. This results in $n \log n$ dependence of mixing time for both quantum and classical limits, as one can infer from (4.5) and (4.7). On the other hand, instantaneous mixing can show essentially different dependence. For instance, suppose there exists a time $t_0$ at which the particle subject to a walk on one-dimensional cycle graph has exactly uniform distribution, i.e. $P_\alpha(t_0) = 1/N$. In this case the distribution of probability in $n$ dimensions becomes $1/N^n$ as follows from (3.11). Therefore one obtains exact inform instantaneous mixing on $n$-dimensional cycle at the time $t = nt_0$. Consequently, the instantaneous mixing time for quantum walks on hyper-cycle structures is linear in $n$, in contrast to non-instantaneous mixing.

In conclusion, we have studied the dynamics of quantum walks on $n$-dimensional hyper-cycle graph structures. Analytical expressions for the density matrix and probability distribution were obtained considering weak (quantum) and strong (classical) decoherence modes. The two regimes were shown to have significant differences in how the walking particle spreads throughout the graph. Upper-bound estimate to mixing time was also obtained for both cases.

We gratefully acknowledge helpful communications with Christino Tamon. This research was supported by the National Science Foundation, grant DMR-0121146.

## Appendix A

Below we demonstrate important steps on simplification of expression (4.2) to (4.3). First, let us introduce $\tilde{P}_{\alpha_j}(t)$, such that

$$P_{\alpha_j}(t) = \frac{1}{N}\left[1 + \tilde{P}_{\alpha_j}(t)\right]. \tag{A.1}$$

In this case, the expression under the summation sign in (4.2) can be expanded as

$$\mathbf{P}_{\boldsymbol{\alpha}}(t_m) - \frac{1}{N^n} = \frac{1}{N^n}\left[\sum_{j=0}^{n-1}\tilde{P}_{\alpha_j}(t_m/n) + \sum_{j<k}^{n-1}\tilde{P}_{\alpha_j}(t_m/n)\tilde{P}_{\alpha_k}(t_m/n) + \ldots + \underbrace{\tilde{P}_{\alpha_0}(t_m/n)\ldots\tilde{P}_{\alpha_{n-1}}(t_m/n)}_{n}\right]. \tag{A.2}$$

We, then, majorize the magnitude of (A.2) as follows

$$\left|\mathbf{P}_{\boldsymbol{\alpha}}(t_m) - \frac{1}{N^n}\right| \leq \frac{1}{N^n}\left[\sum_{j=0}^{n-1}\left|\tilde{P}_{\alpha_j}(t_m/n)\right| + \sum_{j<k}^{n-1}\left|\tilde{P}_{\alpha_j}(t_m/n)\tilde{P}_{\alpha_k}(t_m/n)\right| + \ldots\right] \leq$$

$$\leq \frac{1}{N^n}\left[\sum_{j=0}^{n-1}\left|\tilde{P}_{\alpha_j}(t_m/n)\right| + \sum_{j=0}^{n-1}\left|\tilde{P}_{\alpha_j}(t_m/n)\right|\sum_{k=0}^{n-1}\left|\tilde{P}_{\alpha_k}(t_m/n)\right| + \ldots\right]. \tag{A.3}$$

Furthermore, by defining

$$A_{\boldsymbol{\alpha}}(t_m) \equiv \sum_{j=0}^{n-1}\left|\tilde{P}_{\alpha_j}(t_m/n)\right|, \tag{A.4}$$

one gets the upper-bound for the sum in (4.2)

$$\sum_{\boldsymbol{\alpha}}\left|\mathbf{P}_{\boldsymbol{\alpha}}(t_m) - \frac{1}{N^n}\right| \leq \frac{1}{N^n}\left\{\sum_{\boldsymbol{\alpha}}A_{\boldsymbol{\alpha}}(t_m) + \sum_{\boldsymbol{\alpha}}\left[A_{\boldsymbol{\alpha}}(t_m)\right]^2 + \ldots\right\} \leq$$

$$\leq \frac{1}{N^n}\left\{\sum_{\boldsymbol{\alpha}}A_{\boldsymbol{\alpha}}(t_m) + \left[\sum_{\boldsymbol{\alpha}}A_{\boldsymbol{\alpha}}(t_m)\right]^2 + \ldots\right\} \leq \frac{1}{N^n}\frac{\tilde{A}(t_m)}{1-\tilde{A}(t_m)} \leq \varepsilon \tag{A.5}$$

assuming that $\tilde{A}(t_m) = \sum_{\boldsymbol{\alpha}}A_{\boldsymbol{\alpha}}(t_m) < 1$. From the last inequality of (A.5) we get

$$\tilde{A}(t_m) \leq \frac{N^n\varepsilon}{1+N^n\varepsilon}. \tag{A.6}$$

Expression (A.6) guaranties our assumption. Collecting all the definitions we get

$$N\sum_{j=0}^{n-1}\sum_{\boldsymbol{\alpha}}\left|P_{\alpha_j}(t_m/n) - \frac{1}{N}\right| \leq \frac{N^n\varepsilon}{1+N^n\varepsilon}. \tag{A.7}$$

The summation can be simplified to

$$N\sum_{j=0}^{n-1}N^{n-1}\sum_{\alpha_j}\left|P_{\alpha_j}(t_m/n) - \frac{1}{N}\right| \leq \frac{N^n\varepsilon}{1+N^n\varepsilon}, \tag{4-10}$$

which, taking into account (4.1), finally makes inequality (4.3)

## Appendix B

Let us obtain expression (4.6), i.e. the upper-bound to mixing time for strong decoherence case. To begin with, we substitute the solution for the probabilities based on (2.11) into the left hand-side of (4.2), considering $n = 1$. It yields

$$\sum_{\alpha=0}^{N-1}\left|P_\alpha(t_m)-\frac{1}{N}\right|=\frac{1}{N}\sum_{\alpha=0}^{N-1}\left|\sum_{\mu=1}^{N-1}e^{-\frac{t_m}{\Gamma}\sin^2\left(\frac{\pi\mu}{N}\right)}e^{\frac{2\pi i\mu\alpha}{N}}\right|. \qquad (B.1)$$

We can write the upper-bound to the right hand-side of (B.1) by neglecting the second exponential since it does not exceed unity by the magnitude. This will also simplify the summation

$$\sum_{\alpha=0}^{N-1}\left|P_\alpha(t_m)-\frac{1}{N}\right|\leq\sum_{\mu=1}^{N-1}e^{-\frac{t_m}{\Gamma}\sin^2\left(\frac{\pi\mu}{N}\right)}. \qquad (B.2)$$

At this point, one can notice that the summation over $\mu$ splits into two identical portions. Therefore, one yields

$$\sum_{\alpha=0}^{N-1}\left|P_\alpha(t_m)-\frac{1}{N}\right|\leq\sum_{\mu=1}^{\lfloor N/2\rfloor}e^{-\frac{4\mu t_m}{\Gamma N^2}}. \qquad (B.3)$$

Here, we have utilize the fact that for $0<x<\pi/2$ there is the relation $\sin x>2x/\pi$ and that for $\mu\geq 1$ one has $\mu^2\geq\mu$. Performing the summation in (B.3) we obtain

$$\sum_{\alpha=0}^{N-1}\left|P_\alpha(t_m)-\frac{1}{N}\right|\leq\frac{2}{e^{4t_m/\Gamma N^2}-1}. \qquad (B.4)$$

The right hand-side of (B.4) should now be kept smaller or equal to $\varepsilon$, which finally yields expression (4.6).